\documentclass[aps,prb,twocolumn,showpacs]{revtex4-1}

\usepackage{amsmath}
\usepackage{amssymb}
\usepackage{graphicx}
\usepackage{epstopdf}

\newcommand{\eq}{\mathrm{(0)}}
\newcommand{\fl}{\mathrm{th}}

\begin{document}

\title{Landau-Lifshitz theory of the thermomagnonic torque}

\author{Se Kwon Kim}
\affiliation{
	Department of Physics and Astronomy,
	University of California,
	Los Angeles, California 90095, USA
}

\author{Yaroslav Tserkovnyak}
\affiliation{
	Department of Physics and Astronomy,
	University of California,
	Los Angeles, California 90095, USA
}

\date{\today}

\begin{abstract}
We derive the thermomagnonic torque associated with smooth magnetic textures subjected to a temperature gradient, in the framework of the stochastic Landau-Lifshitz-Gilbert equation. Our approach captures on equal footing two distinct contributions: (1) A local entropic torque that is caused by a temperature dependence of the effective exchange field, the existence of which had been previously suggested based on numerics and (2) the well-known spin-transfer torque induced by thermally-induced magnon flow. The dissipative components of two torques have the same structure, following a common phenomenology, but opposite signs, with the twice larger entropic torque leading to a domain-wall motion toward the hotter region. We compare the efficiency of the torque-driven domain-wall motion with the recently proposed Brownian thermophoresis.
\end{abstract}

\pacs{75.78.-n, 75.30.Ds, 72.20.Pa, 75.10.Hk}

\maketitle

\emph{Introduction.}|Spintronics aims at the active control and manipulation of spin degrees of freedom in condensed-matter systems.\cite{*[][{, and references therein.}] ZuticRMP2004} Spin dynamics can be induced by various means, e.g., an external magnetic field\cite{SchryerJAP1974} or an electric current (in conducting systems, such as magnetic multilayers).\cite{SlonczewskiJMMM1996,*BergerPRB1996} A thermal flux offers another general instrument to excite magnetic dynamics, which forms the central subject of the field dubbed ``spin caloritronics."\cite{*[][{, and references therein.}] BauerNM2012} The spin Seebeck effect, in which a temperature gradient generates spin current, is a prime example of a spin-caloritronic phenomenon, which has been observed in a ferromagnetic metal,\cite{UchidaNature2008} a semiconductor,\cite{JaworskiNM2010} and an insulator.\cite{UchidaNM2010} Thermal control of a magnetic system has an advantage over an electronic control that is necessarily associated with undesired energy dissipation due to electronic continuum. 

A ferromagnetic domain wall is a classic example of topologically-stable magnetic textures; it has been extensively studied due to a fundamental interest as well as practical motivations exemplified by the racetrack memory.\cite{ParkinScience2008} Recent experiments have demonstrated that a temperature gradient drives domain walls to the hotter region.\cite{TorrejonPRL2012,*JiangPRL2013,*ChicoPRB2014} On the theoretical side, a thermomagnonic torque on a domain wall or, more generally, a magnetic texture has been addressed from several perspectives. \textcite{HinzkePRL2011} predicted a thermally-induced motion of a domain wall by numerical simulations based on the stochastic Landau-Lifshitz-Gilbert (LLG) equation;\cite{BrownPR1963, *KuboPTPS1970, *Garcia-PalaciosPRB1998, *ForosPRB2009} \textcite{SchlickeiserPRL2014} argued later that the motion is driven by the so-called entropic torque that originates from the temperature dependence of the exchange stiffness. \textcite{YanPRL2011} derived an adiabatic spin-transfer torque (STT) on the domain wall exerted by magnons based on the conservation of angular momentum: Thermal magnons adjust their spins toward the local magnetic order and react by exerting a torque on the domain wall. \textcite{KovalevEPL2012} constructed a phenomenological theory for the thermomagnonic torque that includes the so-called ``$\beta$-type" dissipative correction to the STT, which has been further elaborated analytically in the framework of the stochastic LLG equation.\cite{KovalevPRB2014}

\begin{figure}[pt]
\includegraphics[width=\columnwidth]{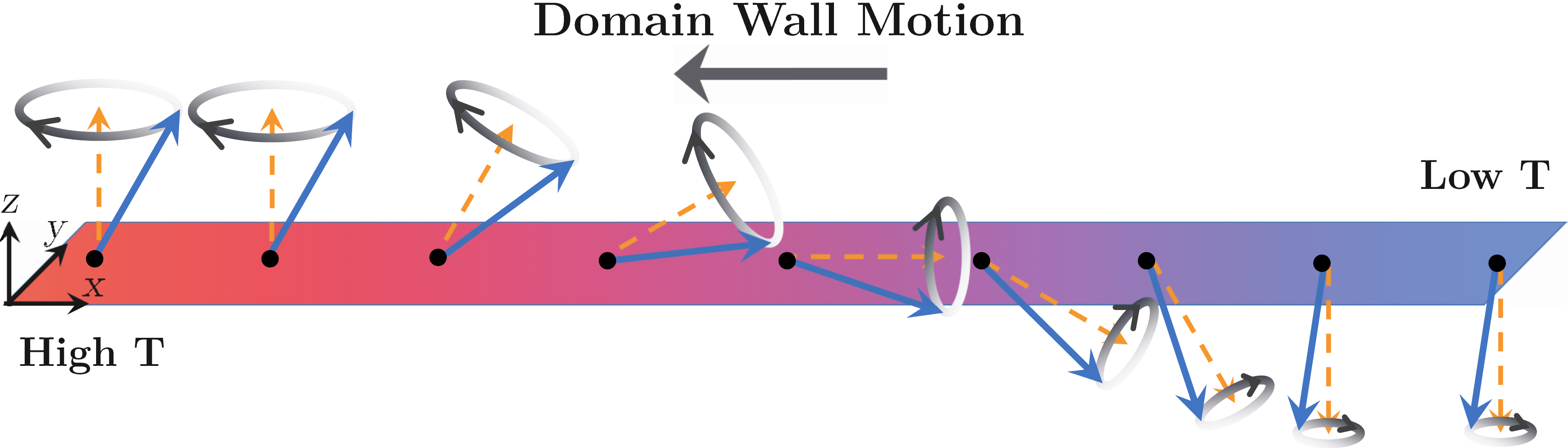}
\caption{(Color online) An illustration of the motion of a domain wall in a one-dimensional ferromagnet under thermal bias. The unit vector $\mathbf{n}$ (blue arrow) in the direction of the local spin density precesses around the equilibrium position due to thermal magnons. As the temperature increases, the length of the time-averaged $\mathbf{n}$ (yellow arrow) decreases, and thus the effective exchange field decreases as well. The temperature dependence of the effective field engenders a torque on the domain wall, pushing it to the hotter region. The STT acting in the opposite direction is not strong enough to reverse the direction of the motion within the LLG treatment.}
\label{fig:dw}
\end{figure}

In this work, we reconcile these theories (resolving some of the conflicting findings) for the thermomagnonic torque within the LLG phenomenology (applicable at temperatures $T\ll T_c$, the Curie temperature), which serves as their common underlying framework. In the following, we start by formulating the model and summarizing our main results: The first is the analytical identification of a contribution to the thermomagnonic torque that is caused by temperature dependence of the effective exchange field, the existence of which had been previously supported by numerics.\cite{SchlickeiserPRL2014} The second is the expression of the total thermomagnonic torque in terms of the magnon current density and the gradient of the magnon-number density, which point to the physical origin of each term. We then provide the detailed derivations of the results. The thermomagnonic torque adds an additional term in the equations of motion for the collective coordinates, which parametrize slow modes of a magnetic texture. We study the motion of a domain wall (sketched in Fig.~\ref{fig:dw}) as an example. Lastly we discuss other possible effects on  thermal motion of magnetic domain walls.

Dynamics of a ferromagnet at finite temperature can be described by the stochastic LLG equation:\cite{BrownPR1963} 
\begin{equation}
s ( 1 + \alpha \mathbf{n} \times ) \dot{\mathbf{n}} = \mathbf{n} \times (\mathbf{h} + \mathbf{h}^\fl),
\label{eq:sllg}
\end{equation}
where $\mathbf{n}$ is the unit vector in the direction of a local spin angular-momentum density $\mathbf{s} \equiv s \mathbf{n}$, $\alpha$ is the Gilbert-damping constant,\cite{GilbertIEEE2004} $\mathbf{h} \equiv - \partial U / \partial \mathbf{n}$ is the effective field conjugate to $\mathbf{n}$. For the ferromagnet's energy, we take, for concreteness, $U[\mathbf{n}] \equiv \int dV [ A |\nabla \mathbf{n}|^2 + K_{ij} n_i n_j ] / 2$, where temperature-independent constants $A > 0$ and $K_{ij}=K_{ji}$ parametrize the exchange stiffness and magnetic anisotropies, respectively. Fluctuations of spins at finite temperature are modeled by the Langevin field $\mathbf{h}^\fl$; the fluctuation-dissipation theorem\cite{ll5} relates the Langevin field to the damping term, which manifests in the correlator of $\mathbf{h}^\fl$ (setting $k_B=1$):
\begin{equation}
\langle h^\fl_i (\mathbf{r},\omega) h^\fl_j (\mathbf{r}',\omega') \rangle =\frac{2\pi\delta_{ij}\alpha s\hbar \omega}{\tanh(\hbar \omega / 2 T)} \delta(\mathbf{r} - \mathbf{r}') \delta(\omega - \omega').
\label{hh}
\end{equation}

\emph{Main results.}|We split $\mathbf{n}$ entering Eq.~(\ref{eq:sllg}) into two orthogonal components: $\mathbf{n} \equiv \sqrt{1 - \delta \mathbf{n}^2} \mathbf{n}^\eq + \delta \mathbf{n}$. The slowly-varying component $\mathbf{n}^\eq$ describes a smooth magnetic texture; the fast component $\delta \mathbf{n}$ represents a small deviation of the local spin density from $\mathbf{n}^\eq$ that is caused by thermal agitation. Thermal magnons constitute the fast component $\delta \mathbf{n}$, whose effect on the magnetic texture is captured by the thermomagnonic torque in the (temporally) coarse-grained LLG equation.

Averaging over the quadratic terms in the fast component $\delta \mathbf{n}$ yields the low-temperature LLG equation (projected transverse to $\mathbf{n}^\eq$) for the slow component $\mathbf{n}^\eq$:
\begin{subequations}
\label{eq:sllg-slow}
\begin{align}
s(1 + \alpha \mathbf{n}^\eq \times) \dot{\mathbf{n}}^\eq 
	= 	& \mathbf{n}^\eq \times \mathbf{h}^\eq \\
	 	& -A \mathbf{n}^\eq \times \partial_i \mathbf{n}^\eq \partial_i \langle\delta \mathbf{n}^2 \rangle \label{eq:torque-1} \\
	 	& +A \langle \delta \mathbf{n} \times \nabla^2 \delta \mathbf{n} \rangle \label{eq:torque-2},
\end{align}
\end{subequations}
in the exchange-magnon approximation, within which only the exchange energy is retained for thermal magnons. Here, $\mathbf{h}^\eq \equiv A \nabla^2 \mathbf{n}^\eq - K_{ij} n^\eq_i \hat{\mathbf{x}}_j$ is the effective field for $\mathbf{n}^\eq$ at zero temperature. \footnote{Note that we are ignoring corrections {$\mathcal{O}(\langle\delta \mathbf{n}^2\rangle)$} to the existing terms, which are insignificant at $T\ll T_c$.}

The second term (\ref{eq:torque-1}) on the right-hand side is a magnonic torque that constitutes our first key result:
\begin{equation}
\boldsymbol{\tau}^\mathrm{ex} \equiv - (2 \hbar A / s) \mathbf{n}^\eq \times (\partial_i \rho \, \partial_i) \mathbf{n}^\eq,
\label{eq:main-1}
\end{equation}
which exists due to temperature dependence of the effective exchange field $A \nabla^2 ( \sqrt{1 - \delta \mathbf{n}^2} \mathbf{n}^\eq )$ through the magnon-number density $\rho \equiv s \langle\delta \mathbf{n}^2\rangle / 2 \hbar$. This contribution is analogous to the so-called entropic torque, which has been argued in Ref.~\onlinecite{SchlickeiserPRL2014} to govern the domain-wall motion.

The third term (\ref{eq:torque-2}), which is identified as the magnonic STT,\cite{YanPRL2011, KovalevEPL2012} can be expressed in terms of the divergence of the spin current $\mathbf{J}^s_i \equiv - A \langle \delta \mathbf{n} \times \partial_i \delta \mathbf{n} \rangle$: $\boldsymbol{\tau}^\mathrm{st}\equiv-\partial_i\mathbf{J}^s_i$. The spin current can be split into the longitudinal and transverse (relative to $\mathbf{n}^\eq$) components:
\begin{equation}
\begin{split}
\mathbf{J}^s_i 	= 	& - A [\mathbf{n}^\eq \cdot \langle \delta \mathbf{n} \times \partial_i \delta \mathbf{n} \rangle] \mathbf{n}^\eq \\
				& - A \mathbf{n}^\eq \times\langle \delta \mathbf{n}(\partial_i \mathbf{n}^\eq \cdot \delta \mathbf{n}) \rangle.
\end{split}
\end{equation}
This finally leads to the following expression for the magnonic STT:
\begin{equation}
\label{eq:tau-stt}
\boldsymbol{\tau}^\mathrm{st} \equiv \hbar (\mathbf{J} \cdot \boldsymbol{\nabla}) \mathbf{n}^\eq + (\hbar A / s) \mathbf{n}^\eq \times (\partial_i \rho \, \partial_i) \mathbf{n}^\eq,
\end{equation}
to first order in the spatial derivative of the texture. Here, $J_i\equiv (A / \hbar) [\mathbf{n}^\eq \cdot \langle \delta \mathbf{n} \times \partial_i \delta \mathbf{n} \rangle]$ is the magnon-flux density evaluated in the absence of a (slow) magnetic texture. For the circular exchange magnons, we similarly evaluated $\langle \delta \mathbf{n}(\partial_i \mathbf{n}^\eq \cdot \delta \mathbf{n}) \rangle=(\hbar\rho/s)\partial_i \mathbf{n}^\eq$.

Adding the two torque contributions, Eqs.~(\ref{eq:main-1}) and (\ref{eq:tau-stt}), yields the total thermomagnonic torque $\boldsymbol{\tau}\equiv\boldsymbol{\tau}^\mathrm{ex}+\boldsymbol{\tau}^\mathrm{st}$:
\begin{equation}
\boldsymbol{\tau} = \hbar (\mathbf{J} \cdot \boldsymbol{\nabla}) \mathbf{n}^\eq - (\hbar A / s) \mathbf{n}^\eq \times (\partial_i \rho \, \partial_i) \mathbf{n}^\eq,
\label{eq:main-2}
\end{equation}
which is a central result of the paper. The benefit of using macroscopic hydrodynamic variables $\mathbf{J}$ and $\boldsymbol{\nabla} \rho$ for the thermomagnonic torque is twofold. First, their directions relative to a temperature gradient are apparent: $\mathbf{J} = - c_J \boldsymbol{\nabla} T$ and $\boldsymbol{\nabla} \rho = c_\rho \boldsymbol{\nabla} T$, with positive constants $c_J$ and $c_\rho$. As a result, the relative sign of two terms in Eq.~(\ref{eq:main-2}) for $\boldsymbol{\tau}$ is unambiguously established, which was not possible in the phenomenological approach.\cite{KovalevEPL2012} Second, it sheds light on the physical origin of the thermally-induced magnetic-texture dynamics. In particular, a ferromagnetic domain wall is driven to the hotter region at the speed $v = \hbar A \partial_i \rho / \alpha s^2$ (normal to the domain-wall orientation), which is governed by the gradient of the magnon-number density. 

The solution for the fast component of the stochastic LLG equation, which we obtain below by the standard approach,\cite{HoffmanPRB2013,KovalevPRB2014} yields in $d$ dimensions:
\begin{equation}
c_J = \frac{I_d}{2^d \pi^{d/2}} \frac{1}{\alpha \hbar \lambda^{d-2}}, \quad c_
\rho = \frac{d I_d}{2^{d+1} \pi^{d/2}} \frac{s}{A\hbar\lambda^{d-2}},
\label{eq:J-rho-soln}
\end{equation}
where $\lambda \equiv \sqrt{\hbar A / s T}$ is the thermal-magnon wavelength. Here, $I_d \equiv [1 / \Gamma(1 + d/2)] \int_0^\infty d\eta \, \eta^{d/2} e^\eta (e^\eta - 1)^{-2}$ is a numerical constant (which would need to be regularized at low energies for $d\leq2$). For $d > 2$, it is given by the Riemann zeta function,\cite{Pathria1996} $I_d = \zeta(d/2)$, with $I_3 \simeq 2.612$ for the physically most relevant case. Note that the first term $\propto\mathbf{J}$ in the thermomagnonic torque $\boldsymbol{\tau}$, Eq.~(\ref{eq:main-2}), becomes dominant in the limit of damping $\alpha\to0$ as then $c_J\to\infty$ while $c_\rho$ remains finite.

The thermomagnonic torque \eqref{eq:main-2} is composed of two terms linear in magnetic texture: The one associated with the magnon-flux density $\mathbf{J}$ is the reactive component of the adiabatic torque.\cite{YanPRL2011, KovalevEPL2012} The other, caused by the gradient of the number density $\rho$, is known as the ``$\beta$-type" dissipative contribution to the adiabatic torque.\cite{KovalevEPL2012,KovalevPRB2014} It has been introduced in the form of $\beta \hbar \mathbf{n}^\eq \times (\mathbf{J} \cdot \boldsymbol{\nabla}) \mathbf{n}^\eq$ by the phenomenological approach.\cite{KovalevEPL2012} Our derivation in the framework of the stochastic LLG equation yields $\beta \mathbf{J}\equiv-(A/s) \boldsymbol{\nabla} \rho$, where $\beta=(A/s)c_\rho/c_J=(d/2)\alpha$.\footnote{It is amusing to observe that Ref.~\onlinecite{KovalevPRB2014} arrived at the same result for $\beta/\alpha$ by overlooking the entropic contribution \eqref{eq:main-1} combined with a sign error in the second term of the STT contribution \eqref{eq:tau-stt}, which together resulted in the correct net torque \eqref{eq:main-2}.} Note that the overall sign of $\beta$ is positive while the STT contribution \eqref{eq:tau-stt} alone would result in a negative $\beta$.\footnote{The sign of $\beta$ depends on the convention for the field $\mathbf{n}$, which is in the direction of the local spin density in our paper. If $\mathbf{n}$ was chosen to be in the direction of the local magnetization of the electronic spin systems with negative gyromagnetic ratio, the sign of $\beta$ would flip.}

\emph{Thermal magnons.}|In order to calculate the magnon flux, $\mathbf{J}$, and number, $\rho$, densities entering in Eq.~\eqref{eq:main-2}, we have to solve the stochastic LLG equation~(\ref{eq:sllg}) for the fast component:
\begin{equation}
	s (\dot{\delta \mathbf{n}} + \alpha \mathbf{n}^\eq \times \dot{\delta \mathbf{n}})
= 	\mathbf{n}^\eq \times A \nabla^2 \delta \mathbf{n}+ \mathbf{n}^\eq \times \mathbf{h}^\fl\,,
\label{eq:sllg-fast}
\end{equation}
to the linear order in $\mathbf{h}^\fl$ and supposing uniform constant order $\mathbf{n}^\eq$. When the average thermal energy of magnons is much larger than the magnon gap,  $T\gg\hbar K/s$, which is usually determined by the strength of the anisotropy $K$, it suffices to retain only the exchange energy of magnons. The fast mode $\delta \mathbf{n}$ can be combined into one complex field $\psi \equiv \delta \mathbf{n} \cdot (\hat{\mathbf{e}}_1 + i \hat{\mathbf{e}}_2)$ within the local orthonormal frame $\{ \hat{\mathbf{e}}_1, \hat{\mathbf{e}}_2, \hat{\mathbf{e}}_3 \equiv \mathbf{n}^\eq \}$. The complex field $\psi$ satisfies the stochastic Schr\"odinger equation with dissipation:\cite{KovalevPRB2014}
\begin{equation}
is (1 + i \alpha) \dot{\psi} = - A \nabla^2 \psi - h.
\label{eq:schrodinger}
\end{equation}
Here the complex Langevin field $h \equiv \mathbf{h}^\fl \cdot (\hat{\mathbf{e}}_1 + i \hat{\mathbf{e}}_2)$ has the correlator, 
$\langle h^* (\mathbf{r}, t) h (\mathbf{r}', t') \rangle = 2\langle \mathbf{h}^\fl (\mathbf{r}, t)\cdot \mathbf{h}^\fl (\mathbf{r}', t') \rangle$  (retaining only the transverse components $\mathbf{h}^\fl\perp\mathbf{n}^\eq$).
The magnon-flux density, $\mathbf{J}$, and the gradient of the magnon-number density, $\boldsymbol{\nabla} \rho$, are given by
\begin{equation}
\mathbf{J} = (A / \hbar) \text{Im}\langle \psi^* \boldsymbol{\nabla} \psi \rangle, \, \boldsymbol{\nabla} \rho = (s / \hbar) \text{Re} \langle \psi^* \boldsymbol{\nabla} \psi \rangle,
\label{Jrho}
\end{equation}
which can be easily obtained from Eq.~(\ref{eq:schrodinger}). Fourier transforming along the transverse coordinates $\mathbf{r}_\perp=(y,z)$ and the time axis,
$
\psi(x, \mathbf{r}_\perp, t) = \int{[d^{d-1} k_\perp d \omega/ (2 \pi)^d] e^{i \mathbf{k}_\perp \cdot \mathbf{r}_\perp-i\omega t} \psi(x, \mathbf{k}_\perp, \omega) },
$
yields the Helmholtz equation:
\begin{equation}
A ( \partial_x^2 + \tilde{k}^2 ) \psi(x, \mathbf{k}_\perp, \omega) = - h(x, \mathbf{k}_\perp, \omega),
\end{equation}
with the complex-valued wave vector
$
\tilde{k}^2 = s \omega (1 + i \alpha) / A - k_\perp^2.
$
The stochastic field correlator is given by 
$
	\langle h^*(x, \mathbf{k}_\perp, \omega) h(x', \mathbf{k}'_\perp, \omega')
\propto 2(2 \pi)^d \delta(\mathbf{k}_\perp - \mathbf{k}'_\perp),
$
with the remaining factors carrying over from Eq.~\eqref{hh}.
Employing the one-dimensional Green's function $G(x - y) = i \exp(i \tilde{k} |x - y|) / 2 \tilde{k}$ [with $\text{Im}(\tilde{k}) > 0$], we finally obtain after straightforward algebra
\begin{equation}
\langle \psi^* (\mathbf{r}, t) \boldsymbol{\nabla} \psi(\mathbf{r}, t) \rangle = \left( \frac{\alpha d}{2} - i \right) \frac{I_d}{2^d \pi^{d/2}} \frac{\boldsymbol{\nabla} T}{\alpha A \lambda^{d-2}},
\label{eq:psi-soln}
\end{equation}
keeping terms that remain in the limit $\alpha \rightarrow 0$. Substituting this result into Eqs.~\eqref{Jrho}, we directly read out coefficients \eqref{eq:J-rho-soln}.

As an additional check, $\boldsymbol{\nabla} \rho$ can be alternatively calculated by considering magnons in the local thermal equilibrium at temperature $T(\mathbf{r})$. Using the Bose-Einstein distribution function, the number density of the thermal exchange magnons is given by $\rho = \int[d^d k/(2 \pi)^d] (e^{\hbar Ak^2 /s T} - 1)^{-1}$, whose gradient reproduces $c_\rho$ in Eq.~(\ref{eq:J-rho-soln}).

\emph{Dynamics of magnetic textures.}|The deterministic magnon-averaged LLG equation for a general smooth texture is given by
\begin{equation}
s(1 + \alpha \mathbf{n}^\eq \times) \dot{\mathbf{n}}^\eq  = \mathbf{n}^\eq \times \mathbf{h}^\eq + \boldsymbol{\tau},
\label{eq:llg}
\end{equation}
in terms of the full thermomagnonic torque $\boldsymbol{\tau} \equiv \boldsymbol{\tau}^\mathrm{ex} + \boldsymbol{\tau}^\mathrm{st}$, Eq.~(\ref{eq:main-2}). The theory can be further simplified by focusing on the relevant collective coordinates parametrizing dynamics of interest. A canonical example of this is a quasi-one-dimensional domain-wall motion described by the azimuthal angle $\Phi(t)$ at the wall's center with position $X(t)$.\cite{SchryerJAP1974} In general, encoding the dynamics of a texture $\mathbf{n}^\eq(\mathbf{r},t)$ with a set of coordinates, $\mathbf{q} (t) = \{ q_1(t), q_2(t), \cdots \}$, the motion of $\mathbf{n}^\eq$ reflects evolution of the coordinates: $\dot{\mathbf{n}}^\eq = \dot{q}_i \partial \mathbf{n}^\eq / \partial q_i$. Taking the inner product of Eq.~(\ref{eq:llg}) with $\mathbf{n}^\eq \times (\partial \mathbf{n}^\eq / \partial q_i)$ followed by spatial integration yields the equations of motion for the collective coordinates:
\begin{equation}
G \dot{\mathbf{q}} - \Gamma \dot{\mathbf{q}} + \mathbf{F} + \mathbf{F}^\mathrm{m} = 0,
\label{eq:eom}
\end{equation}
showing the interplay between gyrotropic, viscous, conservative, and magnon-induced forces acting on a slowly varying texture.\cite{TretiakovPRL2008,*ClarkePRB2008} Here, $G_{ij} \equiv s \int dV \mathbf{n}^\eq \cdot (\partial_{q_j} \mathbf{n}^\eq \times \partial_{q_i} \mathbf{n}^\eq)$ is the antisymmetric gyrotropic tensor; \footnote{If the unit vector $\mathbf{n}$ is in the direction of a local magnetization instead of a local spin angular-momentum density, the sign of $G_{ij}$ would flip} $\Gamma_{ij} \equiv \alpha s \int dV (\partial_{q_i} \mathbf{n}^\eq \cdot \partial_{q_j} \mathbf{n}^\eq)$ is the symmetric viscous tensor; $F_i \equiv - \partial_{q_i} U$ is the conservative force; and, finally,
\begin{equation}
\begin{split}
F^\mathrm{m}_i \equiv & \hbar \int dV [ \mathbf{n}^\eq \times (\partial \mathbf{n}^\eq / \partial q_i) ] \cdot (\mathbf{J} \cdot \boldsymbol{\nabla}) \mathbf{n}^\eq \\
			& - (\hbar A / s) \int dV (\partial \mathbf{n}^\eq/\partial q_i) \cdot (\partial_j \rho\,\partial_j) \mathbf{n}^\eq
\end{split}
\end{equation}
is the thermal-magnon induced force. 

As an illustrative example, we now specialize to the case of a domain-wall motion under the temperature gradient $\boldsymbol{\nabla} T = \partial_x T \hat{\mathbf{x}}$, which is illustrated in Fig.~\ref{fig:dw}. For the easy-$xz$-plane easy-$z$-axis ferromagnet, with energy $U[\mathbf{n}] = \int dV (A |\nabla \mathbf{n}|^2 + K_y n_y^2 - K_z n_z^2) / 2$, $K_y,K_z>0$, a domain wall is a topologically-stable equilibrium defect. It is explicitly given by $\mathbf{n}^\eq = (\sin \theta \cos \Phi, \sin \theta \sin \Phi, \cos \theta)$ with $\Phi=0$ and $\cos \theta = \tanh[(x - X)/\Delta]$, where $\Delta = \sqrt{A/K_z}$ is the width of the wall. Slow dynamics of the domain wall are well described by two collective coordinates: the position $X$ and the azimuthal angle $\Phi$.\cite{SchryerJAP1974} The equations of motion (\ref{eq:eom}) for $X$ and $\Phi$ are
\begin{align}
s (\dot{X} - \alpha \Delta \dot{\Phi}) 	&= - \hbar J_x + (K_y/2) \sin(2 \Phi), \\
s (\Delta \dot{\Phi} + \alpha \dot{X}) 	&= (\hbar A / s) \partial_x \rho.
\end{align}
The steady-state velocity of the wall below the Walker breakdown\cite{SchryerJAP1974} (in the three-dimensional case when wire's lateral dimensions are larger than the thermal-magnon wavelength) is
\begin{equation}
v = \frac{\hbar A}{\alpha s^2} \partial_x \rho = \frac{3 \zeta(3/2)}{16 \pi^{3/2}} \frac{\partial_x T}{\alpha s \lambda}.
\label{eq:V-soln}
\end{equation}
The associated steady-state azimuthal angle is given (at small biases) by $\Phi\approx\zeta(3/2) \partial_x T / 16\pi^{3/2} \alpha \lambda K_y$. For a numerical estimate, taking $s = \hbar / \text{nm}^3$, $\alpha = 10^{-2}$, $\lambda = 10$ nm, and $\partial_x T = 10$ K/mm, the resultant velocity of the wall is $v \simeq 1.15$ mm/s. The domain wall moves to the hotter region, with the driving force proportional to the gradient of the magnon-number density, which is consistent\footnote{Note, however, that for an analytical estimate, \textcite{SchlickeiserPRL2014} approximated the temperature dependence of the exchange stiffness as $d A_\text{eff} / d T \sim A / T_c \sim 1/ 4 a$ (where $a$ is the lattice constant and $A$ is the exchange stiffness at zero temperature) in their Eq.~(10), which appears to yield a gross overestimate at low temperatures (compared to $T_c$). As we have shown, the effective exchange field for long-wavelength dynamics of $\mathbf{n}^{(0)}$ decreases as $\delta A_\text{eff} \sim - |\delta \mathbf{n}|^2 A$ due to thermal magnons. The temperature dependence of $\delta A_\text{eff}$ is thus $d (\delta A_\text{eff}) / d T \sim - 3 I_3 / 8 \pi^{3/2} \lambda$ (in three dimensions), which is smaller than the estimate in Ref.~\onlinecite{SchlickeiserPRL2014} by a factor of $\sim a / \lambda$.} with recent numerical findings.\cite{SchlickeiserPRL2014} 

\emph{Discussion.}|In developing the long-wavelength theory for magnetic-texture dynamics subject to thermal gradients, we have tacitly assumed that the thermal wavelength of magnons is much shorter than the characteristic length scale of the slow texture. This allows us to focus on the adiabatic torques that are first order in spatial-texture gradients. Further simplifications are afforded by assuming energy-scale hierarchy $\hbar K/s\ll T\ll T_c$, such that thermal magnons are dilute and dominated by exchange interactions (hence circular). A strong anisotropy on the atomistic scale, however, would require one to account for the noncircular character of magnons beyond the exchange approximation. Topological defects, such as domain walls, in this case, would, furthermore, be associated with textures that can be sharp compared to the lattice constant. This would lead to strong nonadiabatic effects in magnon transport and torques, engendered by magnon reflection on sharp textures, \cite{YanPRL2012, YanarXiv2015} that are beyond our formalism. High temperatures approaching $T_c$, furthermore, would invalidate the Landau-Lifshitz phenomenology as a starting point, calling for alternative approaches.

Whereas we focused on deterministic texture dynamics in the long-wavelength limit of the bulk, a small topological soliton, e.g., a domain wall in a narrow wire, should generally behave as a particle immersed in a viscous medium. As such, it must exhibit Brownian motion due to random forces required according to the fluctuation-dissipation theorem. We have recently pointed out that a small antiferromagnetic soliton could drift to a colder region by the Brownian motion under a temperature gradient.\cite{KimarXiv2015-2} Naturally, small ferromagnetic solitons behave similarly. For a ferromagnetic domain wall of the type considered above, the drift velocity by the Brownian motion is $v_B = - (\Delta / 2 \alpha s \sigma) \partial_x T$, where $\sigma$ is the magnetic wire's cross section in the $yz$ plane. The ratio between the deterministic torque-induced domain-wall velocity $v$, Eq.~(\ref{eq:V-soln}), and the stochastic Brownian drift $v_B$ is $v / v_B \sim - \sigma / \lambda \Delta$. The thermomagnonic torque thus dominates for larger cross sections $\sigma$ (supposing rigid motion) and/or at higher temperatures, corresponding to shorter thermal-magnon wavelengths $\lambda\propto T^{-1/2}$.

\begin{acknowledgments}
We are grateful for insightful discussions with Oleg Tchernyshyov and Scott Bender. This work was supported by the ARO under Contract No. 911NF-14-1- 0016 and in part by the US DOE- BES under Award No. DE-SC0012190.
\end{acknowledgments}

\bibliography{FDT-FM}

\begin{thebibliography}{35}%
\makeatletter
\providecommand \@ifxundefined [1]{%
 \@ifx{#1\undefined}
}%
\providecommand \@ifnum [1]{%
 \ifnum #1\expandafter \@firstoftwo
 \else \expandafter \@secondoftwo
 \fi
}%
\providecommand \@ifx [1]{%
 \ifx #1\expandafter \@firstoftwo
 \else \expandafter \@secondoftwo
 \fi
}%
\providecommand \natexlab [1]{#1}%
\providecommand \enquote  [1]{``#1''}%
\providecommand \bibnamefont  [1]{#1}%
\providecommand \bibfnamefont [1]{#1}%
\providecommand \citenamefont [1]{#1}%
\providecommand \href@noop [0]{\@secondoftwo}%
\providecommand \href [0]{\begingroup \@sanitize@url \@href}%
\providecommand \@href[1]{\@@startlink{#1}\@@href}%
\providecommand \@@href[1]{\endgroup#1\@@endlink}%
\providecommand \@sanitize@url [0]{\catcode `\\12\catcode `\$12\catcode
  `\&12\catcode `\#12\catcode `\^12\catcode `\_12\catcode `\%12\relax}%
\providecommand \@@startlink[1]{}%
\providecommand \@@endlink[0]{}%
\providecommand \url  [0]{\begingroup\@sanitize@url \@url }%
\providecommand \@url [1]{\endgroup\@href {#1}{\urlprefix }}%
\providecommand \urlprefix  [0]{URL }%
\providecommand \Eprint [0]{\href }%
\providecommand \doibase [0]{http://dx.doi.org/}%
\providecommand \selectlanguage [0]{\@gobble}%
\providecommand \bibinfo  [0]{\@secondoftwo}%
\providecommand \bibfield  [0]{\@secondoftwo}%
\providecommand \translation [1]{[#1]}%
\providecommand \BibitemOpen [0]{}%
\providecommand \bibitemStop [0]{}%
\providecommand \bibitemNoStop [0]{.\EOS\space}%
\providecommand \EOS [0]{\spacefactor3000\relax}%
\providecommand \BibitemShut  [1]{\csname bibitem#1\endcsname}%
\let\auto@bib@innerbib\@empty
\bibitem [{\citenamefont {\ifmmode \check{Z}\else
  \v{Z}\fi{}uti\ifmmode~\acute{c}\else \'{c}\fi{}}\ \emph
  {et~al.}(2004)\citenamefont {\ifmmode \check{Z}\else
  \v{Z}\fi{}uti\ifmmode~\acute{c}\else \'{c}\fi{}}, \citenamefont {Fabian},\
  and\ \citenamefont {Das~Sarma}}]{ZuticRMP2004}%
  \BibitemOpen
  \bibfield  {author} {\bibinfo {author} {\bibfnamefont {I.}~\bibnamefont
  {\ifmmode \check{Z}\else \v{Z}\fi{}uti\ifmmode~\acute{c}\else \'{c}\fi{}}},
  \bibinfo {author} {\bibfnamefont {J.}~\bibnamefont {Fabian}}, \ and\ \bibinfo
  {author} {\bibfnamefont {S.}~\bibnamefont {Das~Sarma}},\ }\href {\doibase
  10.1103/RevModPhys.76.323} {\bibfield  {journal} {\bibinfo  {journal} {Rev.
  Mod. Phys.}\ }\textbf {\bibinfo {volume} {76}},\ \bibinfo {pages} {323}
  (\bibinfo {year} {2004})}\BibitemShut {NoStop}%
\bibitem [{\citenamefont {Schryer}\ and\ \citenamefont
  {Walker}(1974)}]{SchryerJAP1974}%
  \BibitemOpen
  \bibfield  {author} {\bibinfo {author} {\bibfnamefont {N.~L.}\ \bibnamefont
  {Schryer}}\ and\ \bibinfo {author} {\bibfnamefont {L.~R.}\ \bibnamefont
  {Walker}},\ }\href {\doibase http://dx.doi.org/10.1063/1.1663252} {\bibfield
  {journal} {\bibinfo  {journal} {J. Appl. Phys.}\ }\textbf {\bibinfo {volume}
  {45}},\ \bibinfo {pages} {5406} (\bibinfo {year} {1974})}\BibitemShut
  {NoStop}%
\bibitem [{\citenamefont {Slonczewski}(1996)}]{SlonczewskiJMMM1996}%
  \BibitemOpen
  \bibfield  {author} {\bibinfo {author} {\bibfnamefont {J.}~\bibnamefont
  {Slonczewski}},\ }\href {\doibase
  http://dx.doi.org/10.1016/0304-8853(96)00062-5} {\bibfield  {journal}
  {\bibinfo  {journal} {J. Magn. Magn. Mater.}\ }\textbf {\bibinfo {volume}
  {159}},\ \bibinfo {pages} {L1 } (\bibinfo {year} {1996})}\BibitemShut
  {NoStop}%
\bibitem [{\citenamefont {Berger}(1996)}]{BergerPRB1996}%
  \BibitemOpen
  \bibfield  {author} {\bibinfo {author} {\bibfnamefont {L.}~\bibnamefont
  {Berger}},\ }\href {\doibase 10.1103/PhysRevB.54.9353} {\bibfield  {journal}
  {\bibinfo  {journal} {Phys. Rev. B}\ }\textbf {\bibinfo {volume} {54}},\
  \bibinfo {pages} {9353} (\bibinfo {year} {1996})}\BibitemShut {NoStop}%
\bibitem [{\citenamefont {Bauer}\ \emph {et~al.}(2012)\citenamefont {Bauer},
  \citenamefont {Saitoh},\ and\ \citenamefont {van Wees}}]{BauerNM2012}%
  \BibitemOpen
  \bibfield  {author} {\bibinfo {author} {\bibfnamefont {G.~E.~W.}\
  \bibnamefont {Bauer}}, \bibinfo {author} {\bibfnamefont {E.}~\bibnamefont
  {Saitoh}}, \ and\ \bibinfo {author} {\bibfnamefont {B.~J.}\ \bibnamefont {van
  Wees}},\ }\href {http://dx.doi.org/10.1038/nmat3301} {\bibfield  {journal}
  {\bibinfo  {journal} {Nat. Mater.}\ }\textbf {\bibinfo {volume} {11}},\
  \bibinfo {pages} {391} (\bibinfo {year} {2012})}\BibitemShut {NoStop}%
\bibitem [{\citenamefont {Uchida}\ \emph {et~al.}(2008)\citenamefont {Uchida},
  \citenamefont {Takahashi}, \citenamefont {Harii}, \citenamefont {Ieda},
  \citenamefont {Koshibae}, \citenamefont {Ando}, \citenamefont {Maekawa},\
  and\ \citenamefont {Saitoh}}]{UchidaNature2008}%
  \BibitemOpen
  \bibfield  {author} {\bibinfo {author} {\bibfnamefont {K.}~\bibnamefont
  {Uchida}}, \bibinfo {author} {\bibfnamefont {S.}~\bibnamefont {Takahashi}},
  \bibinfo {author} {\bibfnamefont {K.}~\bibnamefont {Harii}}, \bibinfo
  {author} {\bibfnamefont {J.}~\bibnamefont {Ieda}}, \bibinfo {author}
  {\bibfnamefont {W.}~\bibnamefont {Koshibae}}, \bibinfo {author}
  {\bibfnamefont {K.}~\bibnamefont {Ando}}, \bibinfo {author} {\bibfnamefont
  {S.}~\bibnamefont {Maekawa}}, \ and\ \bibinfo {author} {\bibfnamefont
  {E.}~\bibnamefont {Saitoh}},\ }\href {http://dx.doi.org/10.1038/nature07321}
  {\bibfield  {journal} {\bibinfo  {journal} {Nature}\ }\textbf {\bibinfo
  {volume} {455}},\ \bibinfo {pages} {778} (\bibinfo {year}
  {2008})}\BibitemShut {NoStop}%
\bibitem [{\citenamefont {Jaworski}\ \emph {et~al.}(2010)\citenamefont
  {Jaworski}, \citenamefont {Yang}, \citenamefont {Mack}, \citenamefont
  {Awschalom}, \citenamefont {Heremans},\ and\ \citenamefont
  {Myers}}]{JaworskiNM2010}%
  \BibitemOpen
  \bibfield  {author} {\bibinfo {author} {\bibfnamefont {C.~M.}\ \bibnamefont
  {Jaworski}}, \bibinfo {author} {\bibfnamefont {J.}~\bibnamefont {Yang}},
  \bibinfo {author} {\bibfnamefont {S.}~\bibnamefont {Mack}}, \bibinfo {author}
  {\bibfnamefont {D.~D.}\ \bibnamefont {Awschalom}}, \bibinfo {author}
  {\bibfnamefont {J.~P.}\ \bibnamefont {Heremans}}, \ and\ \bibinfo {author}
  {\bibfnamefont {R.~C.}\ \bibnamefont {Myers}},\ }\href
  {http://dx.doi.org/10.1038/nmat2860} {\bibfield  {journal} {\bibinfo
  {journal} {Nat. Mater.}\ }\textbf {\bibinfo {volume} {9}},\ \bibinfo {pages}
  {898} (\bibinfo {year} {2010})}\BibitemShut {NoStop}%
\bibitem [{\citenamefont {Uchida}\ \emph {et~al.}(2010)\citenamefont {Uchida},
  \citenamefont {Xiao}, \citenamefont {Adachi}, \citenamefont {Ohe},
  \citenamefont {Takahashi}, \citenamefont {Ieda}, \citenamefont {Ota},
  \citenamefont {Kajiwara}, \citenamefont {Umezawa}, \citenamefont {Kawai},
  \citenamefont {Bauer}, \citenamefont {Maekawa},\ and\ \citenamefont
  {Saitoh}}]{UchidaNM2010}%
  \BibitemOpen
  \bibfield  {author} {\bibinfo {author} {\bibfnamefont {K.}~\bibnamefont
  {Uchida}}, \bibinfo {author} {\bibfnamefont {J.}~\bibnamefont {Xiao}},
  \bibinfo {author} {\bibfnamefont {H.}~\bibnamefont {Adachi}}, \bibinfo
  {author} {\bibfnamefont {J.}~\bibnamefont {Ohe}}, \bibinfo {author}
  {\bibfnamefont {S.}~\bibnamefont {Takahashi}}, \bibinfo {author}
  {\bibfnamefont {J.}~\bibnamefont {Ieda}}, \bibinfo {author} {\bibfnamefont
  {T.}~\bibnamefont {Ota}}, \bibinfo {author} {\bibfnamefont {Y.}~\bibnamefont
  {Kajiwara}}, \bibinfo {author} {\bibfnamefont {H.}~\bibnamefont {Umezawa}},
  \bibinfo {author} {\bibfnamefont {H.}~\bibnamefont {Kawai}}, \bibinfo
  {author} {\bibfnamefont {G.~E.~W.}\ \bibnamefont {Bauer}}, \bibinfo {author}
  {\bibfnamefont {S.}~\bibnamefont {Maekawa}}, \ and\ \bibinfo {author}
  {\bibfnamefont {E.}~\bibnamefont {Saitoh}},\ }\href
  {http://dx.doi.org/10.1038/nmat2856} {\bibfield  {journal} {\bibinfo
  {journal} {Nat. Mater.}\ }\textbf {\bibinfo {volume} {9}},\ \bibinfo {pages}
  {894} (\bibinfo {year} {2010})}\BibitemShut {NoStop}%
\bibitem [{\citenamefont {Parkin}\ \emph {et~al.}(2008)\citenamefont {Parkin},
  \citenamefont {Hayashi},\ and\ \citenamefont {Thomas}}]{ParkinScience2008}%
  \BibitemOpen
  \bibfield  {author} {\bibinfo {author} {\bibfnamefont {S.~S.~P.}\
  \bibnamefont {Parkin}}, \bibinfo {author} {\bibfnamefont {M.}~\bibnamefont
  {Hayashi}}, \ and\ \bibinfo {author} {\bibfnamefont {L.}~\bibnamefont
  {Thomas}},\ }\href {\doibase 10.1126/science.1145799} {\bibfield  {journal}
  {\bibinfo  {journal} {Science}\ }\textbf {\bibinfo {volume} {320}},\ \bibinfo
  {pages} {190} (\bibinfo {year} {2008})}\BibitemShut {NoStop}%
\bibitem [{\citenamefont {Torrejon}\ \emph {et~al.}(2012)\citenamefont
  {Torrejon}, \citenamefont {Malinowski}, \citenamefont {Pelloux},
  \citenamefont {Weil}, \citenamefont {Thiaville}, \citenamefont {Curiale},
  \citenamefont {Lacour}, \citenamefont {Montaigne},\ and\ \citenamefont
  {Hehn}}]{TorrejonPRL2012}%
  \BibitemOpen
  \bibfield  {author} {\bibinfo {author} {\bibfnamefont {J.}~\bibnamefont
  {Torrejon}}, \bibinfo {author} {\bibfnamefont {G.}~\bibnamefont
  {Malinowski}}, \bibinfo {author} {\bibfnamefont {M.}~\bibnamefont {Pelloux}},
  \bibinfo {author} {\bibfnamefont {R.}~\bibnamefont {Weil}}, \bibinfo {author}
  {\bibfnamefont {A.}~\bibnamefont {Thiaville}}, \bibinfo {author}
  {\bibfnamefont {J.}~\bibnamefont {Curiale}}, \bibinfo {author} {\bibfnamefont
  {D.}~\bibnamefont {Lacour}}, \bibinfo {author} {\bibfnamefont
  {F.}~\bibnamefont {Montaigne}}, \ and\ \bibinfo {author} {\bibfnamefont
  {M.}~\bibnamefont {Hehn}},\ }\href {\doibase 10.1103/PhysRevLett.109.106601}
  {\bibfield  {journal} {\bibinfo  {journal} {Phys. Rev. Lett.}\ }\textbf
  {\bibinfo {volume} {109}},\ \bibinfo {pages} {106601} (\bibinfo {year}
  {2012})}\BibitemShut {NoStop}%
\bibitem [{\citenamefont {Jiang}\ \emph {et~al.}(2013)\citenamefont {Jiang},
  \citenamefont {Upadhyaya}, \citenamefont {Fan}, \citenamefont {Zhao},
  \citenamefont {Wang}, \citenamefont {Chang}, \citenamefont {Lang},
  \citenamefont {Wong}, \citenamefont {Lewis}, \citenamefont {Lin},
  \citenamefont {Tang}, \citenamefont {Cherepov}, \citenamefont {Zhou},
  \citenamefont {Tserkovnyak}, \citenamefont {Schwartz},\ and\ \citenamefont
  {Wang}}]{JiangPRL2013}%
  \BibitemOpen
  \bibfield  {author} {\bibinfo {author} {\bibfnamefont {W.}~\bibnamefont
  {Jiang}}, \bibinfo {author} {\bibfnamefont {P.}~\bibnamefont {Upadhyaya}},
  \bibinfo {author} {\bibfnamefont {Y.}~\bibnamefont {Fan}}, \bibinfo {author}
  {\bibfnamefont {J.}~\bibnamefont {Zhao}}, \bibinfo {author} {\bibfnamefont
  {M.}~\bibnamefont {Wang}}, \bibinfo {author} {\bibfnamefont {L.-T.}\
  \bibnamefont {Chang}}, \bibinfo {author} {\bibfnamefont {M.}~\bibnamefont
  {Lang}}, \bibinfo {author} {\bibfnamefont {K.~L.}\ \bibnamefont {Wong}},
  \bibinfo {author} {\bibfnamefont {M.}~\bibnamefont {Lewis}}, \bibinfo
  {author} {\bibfnamefont {Y.-T.}\ \bibnamefont {Lin}}, \bibinfo {author}
  {\bibfnamefont {J.}~\bibnamefont {Tang}}, \bibinfo {author} {\bibfnamefont
  {S.}~\bibnamefont {Cherepov}}, \bibinfo {author} {\bibfnamefont
  {X.}~\bibnamefont {Zhou}}, \bibinfo {author} {\bibfnamefont {Y.}~\bibnamefont
  {Tserkovnyak}}, \bibinfo {author} {\bibfnamefont {R.~N.}\ \bibnamefont
  {Schwartz}}, \ and\ \bibinfo {author} {\bibfnamefont {K.~L.}\ \bibnamefont
  {Wang}},\ }\href {\doibase 10.1103/PhysRevLett.110.177202} {\bibfield
  {journal} {\bibinfo  {journal} {Phys. Rev. Lett.}\ }\textbf {\bibinfo
  {volume} {110}},\ \bibinfo {pages} {177202} (\bibinfo {year}
  {2013})}\BibitemShut {NoStop}%
\bibitem [{\citenamefont {Chico}\ \emph {et~al.}(2014)\citenamefont {Chico},
  \citenamefont {Etz}, \citenamefont {Bergqvist}, \citenamefont {Eriksson},
  \citenamefont {Fransson}, \citenamefont {Delin},\ and\ \citenamefont
  {Bergman}}]{ChicoPRB2014}%
  \BibitemOpen
  \bibfield  {author} {\bibinfo {author} {\bibfnamefont {J.}~\bibnamefont
  {Chico}}, \bibinfo {author} {\bibfnamefont {C.}~\bibnamefont {Etz}}, \bibinfo
  {author} {\bibfnamefont {L.}~\bibnamefont {Bergqvist}}, \bibinfo {author}
  {\bibfnamefont {O.}~\bibnamefont {Eriksson}}, \bibinfo {author}
  {\bibfnamefont {J.}~\bibnamefont {Fransson}}, \bibinfo {author}
  {\bibfnamefont {A.}~\bibnamefont {Delin}}, \ and\ \bibinfo {author}
  {\bibfnamefont {A.}~\bibnamefont {Bergman}},\ }\href {\doibase
  10.1103/PhysRevB.90.014434} {\bibfield  {journal} {\bibinfo  {journal} {Phys.
  Rev. B}\ }\textbf {\bibinfo {volume} {90}},\ \bibinfo {pages} {014434}
  (\bibinfo {year} {2014})}\BibitemShut {NoStop}%
\bibitem [{\citenamefont {Hinzke}\ and\ \citenamefont
  {Nowak}(2011)}]{HinzkePRL2011}%
  \BibitemOpen
  \bibfield  {author} {\bibinfo {author} {\bibfnamefont {D.}~\bibnamefont
  {Hinzke}}\ and\ \bibinfo {author} {\bibfnamefont {U.}~\bibnamefont {Nowak}},\
  }\href {\doibase 10.1103/PhysRevLett.107.027205} {\bibfield  {journal}
  {\bibinfo  {journal} {Phys. Rev. Lett.}\ }\textbf {\bibinfo {volume} {107}},\
  \bibinfo {pages} {027205} (\bibinfo {year} {2011})}\BibitemShut {NoStop}%
\bibitem [{\citenamefont {Brown}(1963)}]{BrownPR1963}%
  \BibitemOpen
  \bibfield  {author} {\bibinfo {author} {\bibfnamefont {W.~F.}\ \bibnamefont
  {Brown}},\ }\href {\doibase 10.1103/PhysRev.130.1677} {\bibfield  {journal}
  {\bibinfo  {journal} {Phys. Rev.}\ }\textbf {\bibinfo {volume} {130}},\
  \bibinfo {pages} {1677} (\bibinfo {year} {1963})}\BibitemShut {NoStop}%
\bibitem [{\citenamefont {Kubo}\ and\ \citenamefont
  {Hashitsume}(1970)}]{KuboPTPS1970}%
  \BibitemOpen
  \bibfield  {author} {\bibinfo {author} {\bibfnamefont {R.}~\bibnamefont
  {Kubo}}\ and\ \bibinfo {author} {\bibfnamefont {N.}~\bibnamefont
  {Hashitsume}},\ }\href {\doibase 10.1143/PTPS.46.210} {\bibfield  {journal}
  {\bibinfo  {journal} {Prog. Theor. Phys. Suppl.}\ }\textbf {\bibinfo {volume}
  {46}},\ \bibinfo {pages} {210} (\bibinfo {year} {1970})}\BibitemShut
  {NoStop}%
\bibitem [{\citenamefont {Garc\'ia-Palacios}\ and\ \citenamefont
  {L\'azaro}(1998)}]{Garcia-PalaciosPRB1998}%
  \BibitemOpen
  \bibfield  {author} {\bibinfo {author} {\bibfnamefont {J.~L.}\ \bibnamefont
  {Garc\'ia-Palacios}}\ and\ \bibinfo {author} {\bibfnamefont {F.~J.}\
  \bibnamefont {L\'azaro}},\ }\href {\doibase 10.1103/PhysRevB.58.14937}
  {\bibfield  {journal} {\bibinfo  {journal} {Phys. Rev. B}\ }\textbf {\bibinfo
  {volume} {58}},\ \bibinfo {pages} {14937} (\bibinfo {year}
  {1998})}\BibitemShut {NoStop}%
\bibitem [{\citenamefont {Foros}\ \emph {et~al.}(2009)\citenamefont {Foros},
  \citenamefont {Brataas}, \citenamefont {Bauer},\ and\ \citenamefont
  {Tserkovnyak}}]{ForosPRB2009}%
  \BibitemOpen
  \bibfield  {author} {\bibinfo {author} {\bibfnamefont {J.}~\bibnamefont
  {Foros}}, \bibinfo {author} {\bibfnamefont {A.}~\bibnamefont {Brataas}},
  \bibinfo {author} {\bibfnamefont {G.~E.~W.}\ \bibnamefont {Bauer}}, \ and\
  \bibinfo {author} {\bibfnamefont {Y.}~\bibnamefont {Tserkovnyak}},\ }\href
  {\doibase 10.1103/PhysRevB.79.214407} {\bibfield  {journal} {\bibinfo
  {journal} {Phys. Rev. B}\ }\textbf {\bibinfo {volume} {79}},\ \bibinfo
  {pages} {214407} (\bibinfo {year} {2009})}\BibitemShut {NoStop}%
\bibitem [{\citenamefont {Schlickeiser}\ \emph {et~al.}(2014)\citenamefont
  {Schlickeiser}, \citenamefont {Ritzmann}, \citenamefont {Hinzke},\ and\
  \citenamefont {Nowak}}]{SchlickeiserPRL2014}%
  \BibitemOpen
  \bibfield  {author} {\bibinfo {author} {\bibfnamefont {F.}~\bibnamefont
  {Schlickeiser}}, \bibinfo {author} {\bibfnamefont {U.}~\bibnamefont
  {Ritzmann}}, \bibinfo {author} {\bibfnamefont {D.}~\bibnamefont {Hinzke}}, \
  and\ \bibinfo {author} {\bibfnamefont {U.}~\bibnamefont {Nowak}},\ }\href
  {\doibase 10.1103/PhysRevLett.113.097201} {\bibfield  {journal} {\bibinfo
  {journal} {Phys. Rev. Lett.}\ }\textbf {\bibinfo {volume} {113}},\ \bibinfo
  {pages} {097201} (\bibinfo {year} {2014})}\BibitemShut {NoStop}%
\bibitem [{\citenamefont {Yan}\ \emph {et~al.}(2011)\citenamefont {Yan},
  \citenamefont {Wang},\ and\ \citenamefont {Wang}}]{YanPRL2011}%
  \BibitemOpen
  \bibfield  {author} {\bibinfo {author} {\bibfnamefont {P.}~\bibnamefont
  {Yan}}, \bibinfo {author} {\bibfnamefont {X.~S.}\ \bibnamefont {Wang}}, \
  and\ \bibinfo {author} {\bibfnamefont {X.~R.}\ \bibnamefont {Wang}},\ }\href
  {\doibase 10.1103/PhysRevLett.107.177207} {\bibfield  {journal} {\bibinfo
  {journal} {Phys. Rev. Lett.}\ }\textbf {\bibinfo {volume} {107}},\ \bibinfo
  {pages} {177207} (\bibinfo {year} {2011})}\BibitemShut {NoStop}%
\bibitem [{\citenamefont {Kovalev}\ and\ \citenamefont
  {Tserkovnyak}(2012)}]{KovalevEPL2012}%
  \BibitemOpen
  \bibfield  {author} {\bibinfo {author} {\bibfnamefont {A.~A.}\ \bibnamefont
  {Kovalev}}\ and\ \bibinfo {author} {\bibfnamefont {Y.}~\bibnamefont
  {Tserkovnyak}},\ }\href@noop {} {\bibfield  {journal} {\bibinfo  {journal}
  {Europhys. Lett.}\ }\textbf {\bibinfo {volume} {97}},\ \bibinfo {pages}
  {67002} (\bibinfo {year} {2012})}\BibitemShut {NoStop}%
\bibitem [{\citenamefont {Kovalev}(2014)}]{KovalevPRB2014}%
  \BibitemOpen
  \bibfield  {author} {\bibinfo {author} {\bibfnamefont {A.~A.}\ \bibnamefont
  {Kovalev}},\ }\href {\doibase 10.1103/PhysRevB.89.241101} {\bibfield
  {journal} {\bibinfo  {journal} {Phys. Rev. B}\ }\textbf {\bibinfo {volume}
  {89}},\ \bibinfo {pages} {241101} (\bibinfo {year} {2014})}\BibitemShut
  {NoStop}%
\bibitem [{\citenamefont {Gilbert}(2004)}]{GilbertIEEE2004}%
  \BibitemOpen
  \bibfield  {author} {\bibinfo {author} {\bibfnamefont {T.}~\bibnamefont
  {Gilbert}},\ }\href {\doibase 10.1109/TMAG.2004.836740} {\bibfield  {journal}
  {\bibinfo  {journal} {IEEE Trans. Magn.}\ }\textbf {\bibinfo {volume} {40}},\
  \bibinfo {pages} {3443} (\bibinfo {year} {2004})}\BibitemShut {NoStop}%
\bibitem [{\citenamefont {Landau}\ \emph {et~al.}(1980)\citenamefont {Landau},
  \citenamefont {Lifshitz},\ and\ \citenamefont {Pitaevskii}}]{ll5}%
  \BibitemOpen
  \bibfield  {author} {\bibinfo {author} {\bibfnamefont {L.~D.}\ \bibnamefont
  {Landau}}, \bibinfo {author} {\bibfnamefont {E.~M.}\ \bibnamefont
  {Lifshitz}}, \ and\ \bibinfo {author} {\bibfnamefont {L.~P.}\ \bibnamefont
  {Pitaevskii}},\ }\href@noop {} {\emph {\bibinfo {title} {Statistical Physics,
  Part 1}}},\ \bibinfo {edition} {3rd}\ ed.\ (\bibinfo  {publisher} {Pergamon
  Press, New York},\ \bibinfo {year} {1980})\BibitemShut {NoStop}%
\bibitem [{Note1()}]{Note1}%
  \BibitemOpen
  \bibinfo {note} {Note that we are ignoring corrections {$\protect \mathcal
  {O}(\delimiter "426830A \delta \protect \mathbf {n}^2\delimiter "526930B )$}
  to the existing terms, which are insignificant at $T\ll T_c$.}\BibitemShut
  {Stop}%
\bibitem [{\citenamefont {Hoffman}\ \emph {et~al.}(2013)\citenamefont
  {Hoffman}, \citenamefont {Sato},\ and\ \citenamefont
  {Tserkovnyak}}]{HoffmanPRB2013}%
  \BibitemOpen
  \bibfield  {author} {\bibinfo {author} {\bibfnamefont {S.}~\bibnamefont
  {Hoffman}}, \bibinfo {author} {\bibfnamefont {K.}~\bibnamefont {Sato}}, \
  and\ \bibinfo {author} {\bibfnamefont {Y.}~\bibnamefont {Tserkovnyak}},\
  }\href {\doibase 10.1103/PhysRevB.88.064408} {\bibfield  {journal} {\bibinfo
  {journal} {Phys. Rev. B}\ }\textbf {\bibinfo {volume} {88}},\ \bibinfo
  {pages} {064408} (\bibinfo {year} {2013})}\BibitemShut {NoStop}%
\bibitem [{\citenamefont {Pathria}(1996)}]{Pathria1996}%
  \BibitemOpen
  \bibfield  {author} {\bibinfo {author} {\bibfnamefont {R.~K.}\ \bibnamefont
  {Pathria}},\ }\href@noop {} {\emph {\bibinfo {title} {Statistical
  Mechanics}}},\ \bibinfo {edition} {2nd}\ ed.\ (\bibinfo  {publisher}
  {Butterworth Heinemann, Woburn},\ \bibinfo {year} {1996})\BibitemShut
  {NoStop}%
\bibitem [{Note2()}]{Note2}%
  \BibitemOpen
  \bibinfo {note} {It is amusing to observe that Ref.~\protect \rev@citealpnum
  {KovalevPRB2014} arrived at the same result for $\beta /\alpha $ by
  overlooking the entropic contribution \protect \textup {\hbox {\mathsurround
  \z@ \protect \normalfont (\ignorespaces \ref {eq:main-1}\unskip \@@italiccorr
  )}} combined with a sign error in the second term of the STT contribution
  \protect \textup {\hbox {\mathsurround \z@ \protect \normalfont
  (\ignorespaces \ref {eq:tau-stt}\unskip \@@italiccorr )}}, which together
  resulted in the correct net torque \protect \textup {\hbox {\mathsurround \z@
  \protect \normalfont (\ignorespaces \ref {eq:main-2}\unskip \@@italiccorr
  )}}.}\BibitemShut {Stop}%
\bibitem [{Note3()}]{Note3}%
  \BibitemOpen
  \bibinfo {note} {The sign of $\beta $ depends on the convention for the field
  $\protect \mathbf {n}$, which is in the direction of the local spin density
  in our paper. If $\protect \mathbf {n}$ was chosen to be in the direction of
  the local magnetization of the electronic spin systems with negative
  gyromagnetic ratio, the sign of $\beta $ would flip.}\BibitemShut {Stop}%
\bibitem [{\citenamefont {Tretiakov}\ \emph {et~al.}(2008)\citenamefont
  {Tretiakov}, \citenamefont {Clarke}, \citenamefont {Chern}, \citenamefont
  {Bazaliy},\ and\ \citenamefont {Tchernyshyov}}]{TretiakovPRL2008}%
  \BibitemOpen
  \bibfield  {author} {\bibinfo {author} {\bibfnamefont {O.~A.}\ \bibnamefont
  {Tretiakov}}, \bibinfo {author} {\bibfnamefont {D.}~\bibnamefont {Clarke}},
  \bibinfo {author} {\bibfnamefont {G.-W.}\ \bibnamefont {Chern}}, \bibinfo
  {author} {\bibfnamefont {Y.~B.}\ \bibnamefont {Bazaliy}}, \ and\ \bibinfo
  {author} {\bibfnamefont {O.}~\bibnamefont {Tchernyshyov}},\ }\href {\doibase
  10.1103/PhysRevLett.100.127204} {\bibfield  {journal} {\bibinfo  {journal}
  {Phys. Rev. Lett.}\ }\textbf {\bibinfo {volume} {100}},\ \bibinfo {pages}
  {127204} (\bibinfo {year} {2008})}\BibitemShut {NoStop}%
\bibitem [{\citenamefont {Clarke}\ \emph {et~al.}(2008)\citenamefont {Clarke},
  \citenamefont {Tretiakov}, \citenamefont {Chern}, \citenamefont {Bazaliy},\
  and\ \citenamefont {Tchernyshyov}}]{ClarkePRB2008}%
  \BibitemOpen
  \bibfield  {author} {\bibinfo {author} {\bibfnamefont {D.~J.}\ \bibnamefont
  {Clarke}}, \bibinfo {author} {\bibfnamefont {O.~A.}\ \bibnamefont
  {Tretiakov}}, \bibinfo {author} {\bibfnamefont {G.-W.}\ \bibnamefont
  {Chern}}, \bibinfo {author} {\bibfnamefont {Y.~B.}\ \bibnamefont {Bazaliy}},
  \ and\ \bibinfo {author} {\bibfnamefont {O.}~\bibnamefont {Tchernyshyov}},\
  }\href {\doibase 10.1103/PhysRevB.78.134412} {\bibfield  {journal} {\bibinfo
  {journal} {Phys. Rev. B}\ }\textbf {\bibinfo {volume} {78}},\ \bibinfo
  {pages} {134412} (\bibinfo {year} {2008})}\BibitemShut {NoStop}%
\bibitem [{Note4()}]{Note4}%
  \BibitemOpen
  \bibinfo {note} {If the unit vector $\protect \mathbf {n}$ is in the
  direction of a local magnetization instead of a local spin angular-momentum
  density, the sign of $G_{ij}$ would flip}\BibitemShut {NoStop}%
\bibitem [{Note5()}]{Note5}%
  \BibitemOpen
  \bibinfo {note} {Note, however, that for an analytical estimate, \protect
  \rev@citet {SchlickeiserPRL2014} approximated the temperature dependence of
  the exchange stiffness as $d A_\protect \text {eff} / d T \sim A / T_c \sim
  1/ 4 a$ (where $a$ is the lattice constant and $A$ is the exchange stiffness
  at zero temperature) in their Eq.~(10), which appears to yield a gross
  overestimate at low temperatures (compared to $T_c$). As we have shown, the
  effective exchange field for long-wavelength dynamics of $\protect \mathbf
  {n}^{(0)}$ decreases as $\delta A_\protect \text {eff} \sim - |\delta
  \protect \mathbf {n}|^2 A$ due to thermal magnons. The temperature dependence
  of $\delta A_\protect \text {eff}$ is thus $d (\delta A_\protect \text {eff})
  / d T \sim - 3 I_3 / 8 \pi ^{3/2} \lambda $ (in three dimensions), which is
  smaller than the estimate in Ref.~\protect \rev@citealpnum
  {SchlickeiserPRL2014} by a factor of $\sim a / \lambda $.}\BibitemShut
  {Stop}%
\bibitem [{\citenamefont {Yan}\ and\ \citenamefont {Bauer}(2012)}]{YanPRL2012}%
  \BibitemOpen
  \bibfield  {author} {\bibinfo {author} {\bibfnamefont {P.}~\bibnamefont
  {Yan}}\ and\ \bibinfo {author} {\bibfnamefont {G.~E.~W.}\ \bibnamefont
  {Bauer}},\ }\href {\doibase 10.1103/PhysRevLett.109.087202} {\bibfield
  {journal} {\bibinfo  {journal} {Phys. Rev. Lett.}\ }\textbf {\bibinfo
  {volume} {109}},\ \bibinfo {pages} {087202} (\bibinfo {year}
  {2012})}\BibitemShut {NoStop}%
\bibitem [{\citenamefont {Yan}\ \emph {et~al.}()\citenamefont {Yan},
  \citenamefont {Cao},\ and\ \citenamefont {Sinova}}]{YanarXiv2015}%
  \BibitemOpen
  \bibfield  {author} {\bibinfo {author} {\bibfnamefont {P.}~\bibnamefont
  {Yan}}, \bibinfo {author} {\bibfnamefont {Y.}~\bibnamefont {Cao}}, \ and\
  \bibinfo {author} {\bibfnamefont {J.}~\bibnamefont {Sinova}},\ }\href@noop {}
  {}\Eprint {http://arxiv.org/abs/1504.00651} {arXiv:1504.00651} \BibitemShut
  {NoStop}%
\bibitem [{\citenamefont {Kim}\ \emph {et~al.}()\citenamefont {Kim},
  \citenamefont {Tchernyshyov},\ and\ \citenamefont
  {Tserkovnyak}}]{KimarXiv2015-2}%
  \BibitemOpen
  \bibfield  {author} {\bibinfo {author} {\bibfnamefont {S.~K.}\ \bibnamefont
  {Kim}}, \bibinfo {author} {\bibfnamefont {O.}~\bibnamefont {Tchernyshyov}}, \
  and\ \bibinfo {author} {\bibfnamefont {Y.}~\bibnamefont {Tserkovnyak}},\
  }\href@noop {} {}\Eprint {http://arxiv.org/abs/1503.07854} {arXiv:1503.07854}
  \BibitemShut {NoStop}%
\end{thebibliography}%

\end{document}